\documentclass[prd,onecolumn,nofootinbib,showkeys]{revtex4}
\usepackage{bm,amsmath,amssymb,graphicx}

\begin{document}

\title{NONCOMMUTATIVE MOMENTUM AND TORSIONAL REGULARIZATION}
\author{Nikodem Pop{\l}awski}
\altaffiliation{NPoplawski@newhaven.edu}
\affiliation{Department of Mathematics and Physics, University of New Haven, 300 Boston Post Road, West Haven, CT 06516, USA}

\noindent
{\em Foundations of Physics}\\
Vol. {\bf 50}, No. 9, pp. 900--923 (2020)
\vspace{0.4in}

\begin{abstract}
We show that in the presence of the torsion tensor $S^k_{\phantom{k}ij}$, the quantum commutation relation for the four-momentum, traced over spinor indices, is given by $[p_i,p_j]=2i\hbar S^k_{\phantom{k}ij}p_k$.
In the Einstein--Cartan theory of gravity, in which torsion is coupled to spin of fermions, this relation in a coordinate frame reduces to a commutation relation of noncommutative momentum space, $[p_i,p_j]=i\epsilon_{ijk}Up^3 p_k$, where $U$ is a constant on the order of the squared inverse of the Planck mass.
We propose that this relation replaces the integration in the momentum space in Feynman diagrams with the summation over the discrete momentum eigenvalues.
We derive a prescription for this summation that agrees with convergent integrals:
\[
\int\frac{d^4p}{(p^2+\Delta)^s}\rightarrow 4\pi U^{s-2}\sum_{l=1}^\infty \int_0^{\pi/2} d\phi \frac{\sin^4\phi\,n^{s-3}}{[\sin\phi+U\Delta n]^s},
\]
where $n=\sqrt{l(l+1)}$ and $\Delta$ does not depend on $p$.
We show that this prescription regularizes ultraviolet-divergent integrals in loop diagrams.
We extend this prescription to tensor integrals.
We derive a finite, gauge-invariant vacuum polarization tensor and a finite running coupling.
Including loops from all charged fermions, we find a finite value for the bare electric charge of an electron: $\approx -1.22\,e$.
This torsional regularization may therefore provide a realistic, physical mechanism for eliminating infinities in quantum field theory and making renormalization finite.
\end{abstract}
\maketitle

\section{Introduction}

In quantum electrodynamics (QED) \cite{qft}, a calculation of the amplitude for a physical process must include perturbative corrections involving Feynman diagrams \cite{Feyn} with closed loops of virtual particles (radiative corrections).
The integration in the resulting integrals is taken in the four-momentum space, with the magnitudes of the energy and momentum running to infinity and not restricted to the relativistic energy-momentum relation (off-shell particles).
Many integrals that appear in radiative corrections are divergent, which is referred to as the ultraviolet divergence.
The ultraviolet divergence results from the asymptotic, high-energy behavior of the Feynman propagators \cite{Feyn}.
Physically, the divergence is a consequence of an incompleteness of our understanding of the physics at large energies and momenta \cite{Tom}.
There are three one-loop divergent diagrams in QED: vacuum polarization (a photon creating a virtual electron-positron pair which then annihilates), self-energy (an electron emits and reabsorbs a virtual photon), and vertex (an electron emits a photon, emits a second photon, and then reabsorbs the first) \cite{Schw}.
More complex diagrams can be reduced to the combinations of these three diagrams.
The gauge-invariant parts of all three diagrams are logarithmically divergent.

Mathematically, these divergences can be treated by a method of regularization \cite{qft}.
In Pauli--Villars regularization \cite{PaVi}, the vertices are retained but Feynman propagators are modified.
From a divergent integral involving a particle of mass $m$ one subtracts an integral of the same form but with a different mass $\Lambda$ representing a fictitious particle, and then one takes a limit $\Lambda\rightarrow\infty$.
The divergent term in the difference has a form $\textrm{const}\cdot\textrm{ln}\Lambda$ and is absorbed through redefining the original (bare) mass, charge and field, leaving a finite, physical (dressed) value that is measured in experiment.
Such redefinition is referred to as renormalization \cite{qft,Dys,Gel}.
In 't Hooft--Veltman dimensional regularization \cite{HoVe}, the number of dimensions $n=4$ is replaced by a fictitious number $n=4-\epsilon$, and then one takes a limit $\epsilon\rightarrow 0$.
The divergent term has a form $\textrm{const}/\epsilon$ and is absorbing through renormalization, leaving a finite, physical value.
Dirac was persistently critical about renormalization and expected a realistic regularization based on the established principles of physics \cite{Dir}.

An interesting idea for such a regularization was explored in \cite{Rus}, where ultraviolet divergences in quantum field theory might be avoided by curving momentum space.
The idea that momentum space might be curved was first suggested by Born \cite{Born}.
Curved phase space may have several novel consequences on the motion of particles \cite{Cas}.
Snyder pointed out that the curvature of momentum space implies the noncommutativity of spacetime coordinates \cite{Snyd}.
This observation has led to the development of noncommutative geometry, most notably by Connes \cite{Con}, and then to noncommutative field theory \cite{DoNe}.
Quantum geometry in which a curved momentum space is dual to a noncommutative spacetime was explored in \cite{Maj}.

The theory of relativity postulates that spacetime is the invariant arena for nonquantum physics.
A novel principle of relative locality was suggested in \cite{reloc}, according to which a phase space is the invariant area for nonquantum physics.
In this scenario, both coordinate space and momentum space are curved.
Propagating and interacting particles are observed in spacetime constructed by observers, but observers at different locations construct different spacetime projections from the invariant phase space.
The curvature, torsion, and nonmetricity of momentum space can manifest themselves in various deformations of the
additivity of the momentum and energy, modifying the energy-momentum conservation laws \cite{reloc}.
In addition, curved momentum space is related to the invariant length scale \cite{inv}.

If a curved momentum space implies a noncommutative coordinate spacetime, then a curved coordinate spacetime should imply a noncommutative momentum space.
Quantum mechanics with noncommutative momentum was explored in \cite{nonco}.
The integration in noncommutative momentum space must be replaced with the summation over all eigenvalues of the momentum \cite{eigen}.

In this article, we postulate that a realistic regularization, expected by Dirac, may come from the noncommutativity of momentum.
Moreover, we argue that the noncommutative momentum is a consequence of spacetime torsion.
The existence of torsion \cite{Schr,Schou} is required by the consistency of the conservation law for the total (orbital plus spin) angular momentum of a Dirac particle curved spacetime with relativistic quantum mechanics \cite{req}.
Therefore, we refer to this postulate as {\em torsional regularization}.
We propose the commutation relation for the momentum that is consistent with the Einstein--Cartan (EC) theory of gravity, in which torsion couples to spin \cite{EC}.
The momentum operators do not commute (uncertainty principle for momentum), which becomes significant at larger momenta where regularization is needed.

We propose a prescription for the summation in noncommutative momentum space (replacing the integration in commutative momentum space) that gives a correct continuous limit for convergent integrals.
We show that ultraviolet-divergent integrals turn into convergent sums, naturally eliminating the ultraviolet divergence in loop diagrams.
Using the four-dimensional Gauss theorem, we extend this prescription to tensor integrals.
Then we apply our prescription to vacuum polarization and derive a finite, gauge-invariant vacuum polarization tensor.
We derive a finite running coupling that agrees with the low-energy limit of the standard quantum electrodynamics.
Finally, we find that the modification of the photon propagator arising from the loops involving all charged fermions give a finite value for the bare electric charge of a particle: it is approximately 1.22 times its measured, renormalized charge.
The renormalization paradigm with torsion is finite and thus mathematically self-consistent.
Therefore, torsion may be the source of a realistic regularization which was believed to exist by Dirac, rendering quantum electrodynamics ultraviolet complete.

\section{Torsion and commutation relation for momentum}

In the general theory of relativity (GR), the orbital angular momentum of a free particle is conserved \cite{LL2,Lord,Niko}.
However, according to the Dirac equation in relativistic quantum mechanics, the orbital angular momentum alone is not conserved.
Instead, the sum of the orbital angular momentum and the intrinsic angular momentum (spin) is conserved \cite{Direq}.
Consequently, GR must be extended to allow the exchange between the orbital and spin parts of the angular momentum that occurs in quantum mechanics.
Spacetime with torsion, proposed by Cartan \cite{Cartan}, naturally provides such an extension \cite{req,Lord,Niko}.

Furthermore, torsion may prevent the formation of gravitational singularities.
The simplest and most natural theory of gravity with torsion is the EC theory \cite{EC,Lord,Niko}, in which torsion is generated by the spin of fermions.
EC agrees with solar system, binary pulsar, and cosmological tests of general relativity because even at nuclear densities, the corrections from torsion to the Einstein equations are negligible \cite{EC,Lord}.
The spin-torsion coupling in EC generates gravitational repulsion at extremely high densities and thus avoids the formation of singularities in black holes and at the big bang \cite{avert}.
The collapsing matter in a black hole bounces at a finite density and then expands into a new, finite region of space with positive curvature on the other side of the event horizon, which may be regarded as a new universe \cite{BH}.
Quantum particle production caused by an extremely high curvature near a bounce (which replaces the big bang) creates enormous amounts of matter and entropy, and generates a finite period of exponential expansion (inflation) of this universe \cite{ApJ}.
EC also modifies the Dirac equation, adding a term that is cubic in spinor fields \cite{nonl}.
We proposed in \cite{nons} that this term may solve the problem of divergent integrals in quantum field theory by providing fermions with spatial extension (about $10^{-27}$ m for an electron) and thus introducing an effective ultraviolet cutoff for their propagators.
In this article, we explore that proposal.

The torsion tensor is the antisymmetric part of the affine connection $\Gamma^{\,\,i}_{j\,k}$: $S^i_{\phantom{i}jk}=(1/2)(\Gamma^{\,\,i}_{jk}-\Gamma^{\,\,i}_{kj})$ \cite{Schr,Schou}.
In the presence of torsion, the parallel transports (which define the covariant derivative) do not commute, which results from the following construction \cite{Schou}.
The parallel transport of an infinitesimal, four-dimensional vector $PR=dx^i$ from a point $P$ to an infinitesimally close point $Q$ such that $PQ=dx'^j$ adds to $dx^i$ a small correction: $\delta dx^i = -\Gamma^{\,\,\,i}_{jk}dx^j dx'^k$ \cite{LL2}.
After effecting the transport, the vector $dx^i+\delta dx^i$ points to a point $T$.
The parallel transport of the vector $dx'^i$ from a point $P$ to an infinitesimally close point $R$ adds to $dx^i$ a small correction: $\delta dx'^i = -\Gamma^{\,\,\,i}_{jk}dx^j dx'^k$.
After effecting the transport, the vector $dx'^i+\delta dx'^i$ points to a point $T'$.

Without torsion, points $T$ and $T'$ would coincide and form, together with points $P$, $Q$, and $R$, a parallelogram because $\delta dx'^i-\delta dx^i=\Gamma^{\,\,\,i}_{kj}dx^j dx'^k - \Gamma^{\,\,\,i}_{jk}dx^j dx'^k = 0$.
If the torsion tensor is not zero, however, the affine connection is asymmetric in the lower indices and $\delta dx'^i-\delta dx^i = -2S^i_{\phantom{i}jk}dx^j dx'^k$.
Points $T$ and $T'$ do not coincide, the parallogram is not closed, and the combination of two displacements of point $P$ (through $dx^i$ and $dx'^j$) depends on their order.
Accordingly, covariant derivatives of a scalar $\psi$ do not commute:
\begin{equation}
\nabla_i\nabla_j\psi-\nabla_j\nabla_i\psi=\partial_i\nabla_j\psi-\Gamma^{\,k}_{ji}\nabla_k\psi-\partial_j\nabla_i\psi+\Gamma^{\,k}_{ij}\nabla_k\psi=\partial_i\partial_j\psi-\partial_j\partial_i\psi+2S^k_{\phantom{k}ij}\nabla_k\psi=2S^k_{\phantom{k}ij}\nabla_k\psi.
\label{tors}
\end{equation}

The momentum is defined in mechanics as a generator of a translation \cite{LL1} and the translation is described in terms of the covariant derivative and parallel transport.
The covariant four-dimensional momentum operator in position space is therefore related to the covariant derivative: $p_k=i\hbar\nabla_k$.
This relation generalizes the standard relations $E=i\hbar\partial/\partial t$ and $p_x=-i\hbar\partial/\partial x$.
Putting the covariant momentum operator in (\ref{tors}) gives a
covariant commutation relation for the components of the four-dimensional momentum operator when they act on a scalar wave function:
\begin{equation}
[p_i,p_j]=2i\hbar S^k_{\phantom{k}ij}p_k.
\label{twist}
\end{equation}
This equation indicates that the four-dimensional momentum operators in spacetime with torsion do not commute.

In relativistic quantum mechanics, the wave function is a Dirac spinor \cite{qft}.
The commutator of covariant derivatives of a spinor $\psi$ is given by
\begin{equation}
\nabla_i\nabla_j\psi-\nabla_j\nabla_i\psi=2S^k_{\phantom{k}ij}\nabla_k\psi+\frac{1}{4}R_{klij}\gamma^k \gamma^l \psi,
\label{relativ}
\end{equation}
where $\gamma^k$ are the Dirac matrices in curved spacetime: $\gamma^i\gamma^j+\gamma^j\gamma^i=2g^{ij}I_4$ and $I_4$ is the unit 4$\times$4 matrix \cite{Lord,Niko}.
This commutator differs from (\ref{tors}) by the term with the curvature tensor $R_{ijkl}$.
Putting the covariant momentum operator in (\ref{relativ}) gives a commutation relation for the four-dimensional momentum operators when they act on a spinor wave function:
\begin{equation}
[p_i,p_j]=2i\hbar S^k_{\phantom{k}ij}p_k I_4+\frac{1}{4}i\hbar R_{klij}\gamma^k \gamma^l.
\end{equation}
When this relation is averaged (traced) over spinor indices, which is done in calculations involving Feynman diagrams with closed loops, it reduces to the relation (\ref{twist}) because $R_{klij}g^{kl}=0$.\footnote{
The contraction of the curvature tensor in its first two indices vanishes if spacetime satisfies the metric compatibility of the affine connection: $\nabla_i g_{jk}=0$.
This condition holds in GR and EC.
}
Consequently, the noncommutativity of the momentum four-vector will modify the integration in traced Feynman loops only through torsion, not curvature, and the relation (\ref{twist}) holds.

\section{Einstein--Cartan gravity}

In the EC theory of gravity, torsion is coupled to the spin of fermions and fermions are the source of torsion \cite{EC,Lord,Niko}.
The Lagrangian density for the gravitational field has the same form as that in general relativity: $\mathcal{L}_g=-R\sqrt{-g}/(2\kappa)$, where $R$ is the Ricci scalar constructed from the metric-compatible affine connection, $g$ is the determinant of the metric tensor $g_{ij}$, and $\kappa=8\pi G$ (we use units such that $c=1$).
The variation of the total action for the gravitational field and matter with respect to the metric tensor gives the Einstein field equations.
The variation of the action with respect to the torsion tensor gives the Cartan equations that relate the spin density and torsion:
\begin{equation}
S^i_{\phantom{i}jk}-S_j\delta^i_k+S_k\delta^i_j=-\frac{1}{2}\kappa s_{jk}^{\phantom{jk}i},
\label{Cartan}
\end{equation}
where $S_i=S^k_{\phantom{k}ik}$ is the torsion vector and $s_{ijk}=2\delta\mathcal{L}_m/\delta C^{ijk}$ is the spin tensor of matter (the variational derivative of the Lagrangian density for matter $\mathcal{L}_m$ with respect to the contortion tensor $C_{ijk}=S_{ijk}+S_{jki}+S_{kji}$) (we use the notation of \cite{Niko}).

For the Dirac fields, the spin tensor is completely antisymmetric \cite{EC,Lord,Niko}.
Therefore, the field equations (\ref{Cartan}) determine that the torsion tensor is also completely antisymmetric and thus can be represented as dual to the torsion pseudovector $S^l$: $S_{ijk}=-\epsilon_{ijkl}S^l$, where $\epsilon^{ijkl}$ is the Levi-Civita pseudotensor \cite{LL2}.
The commutation relation (\ref{twist}) becomes $[p_i,p_j]=-2i\hbar\epsilon_{ij\phantom{kl}}^{\phantom{ij}kl}S_l p_k$.
This covariant relation, in a coordinate frame in which the torsion pseudovector has only the time component $S^0=-Q/(2\hbar)$, reduces to a form that includes only the spatial components $p_\alpha$:
\begin{equation}
[p_\alpha,p_\beta]=i\epsilon_{\alpha\beta\gamma}Qp_\gamma,
\label{structure}
\end{equation}
where $\epsilon_{\alpha\beta\gamma}$ is the spatial, totally antisymmetric unit pseudotensor \cite{LL2}.
Torsion therefore provides a noncommutative momentum space that can be realized by the momenta operators satisfying a nonzero commutation relation.
The relation (\ref{structure}) can be written in terms of the components $p_x,p_y,p_z$ of the momentum ${\bf p}$:
\begin{equation}
[p_x,p_y] = iQp_z,\quad [p_y,p_z] = iQp_x,\quad [p_z,p_x] = iQp_y,
\label{com}
\end{equation}
where $Q$ is a function of operators in the phase space.
The momentum components commute with the square of the momentum, ${\bf p}^2={\bf p}\cdot{\bf p}$.
They commute with the energy component $p^0$, from which it follows that ${\bf p}^2$ and $p^0$ commute too.

The general form of the function $Q$ can be determined from the Jacobi identities that must be satisfied by the operators in (\ref{com}) \cite{qm}.
A similar determination for the position operators in noncommutative coordinate space was found in \cite{Mag} and \cite{Jac}.
Analogously to the analysis in \cite{Jac}, which considered a three-dimensional noncommutative coordinate space with a commutation relation $[x^i,x^j]=iC\epsilon_{ijk}x_k$ with a constant $C$, we conclude that $Q$ is a function of ${\bf p}$ and $[x^i,x^j]=0$.
In addition, $[x^i,p^j]$ does not vanish but depends on $x^k$, the metric tensor, and $\epsilon_{ijk}$.
To maintain covariance, the quantity $Q$ must be a function of $p^2=(p^0)^2-{\bf p}^2$, where $p^0$ is the time component of the momentum four-vector $p^\mu=(p^0,{\bf p})$ in a locally flat coordinate frame \cite{LL2}.

The spin tensor on the right-hand side of (\ref{Cartan}) has dimension mass$^3$, and $\kappa$ in the units such that $\hbar=1$ is on the order of $1/m^2_\textrm{P}$, where $m_\textrm{P}$ is the Planck mass.
Therefore, the quantity $Q$ should have dimension mass$^3/m^2_\textrm{P}$.
Since $Q(p)$ in (\ref{structure}) must be a scalar, it should be a cubic function of the magnitude of the momentum four-vector $p$:
\begin{equation}
Q=Up^3,
\label{val}
\end{equation}
where $U$ is a constant on the order of $1/m^2_\textrm{P}$.

Since torsion is significant only at extremely high densities or energies, the right-hand side of (\ref{twist}) is nearly zero at the scales currently available to experiment or observation, effectively reproducing the standard commutation relation $[p_i,p_j]=0$.
Therefore, torsional regularization does not introduce significant deviations in quantum field theory at those scales.

Equations (\ref{com}) resemble the commutation relations for the angular momentum: $[L_x,L_y]=i\hbar L_z$, $[L_y,L_z]=i\hbar L_x$, $[L_z,L_x]=i\hbar L_y$ \cite{qm}.
Those relations derive the separation between adjacent eigenvalues of $L_z$; that separation is $\hbar$.
Analogously, the separation between adjacent eigenvalues of $p_z$ is $|Q|$.
Since $Q\sim p^3$, the separation between the eigenvalues of the momentum increases as the magnitude of the momentum vector increases.
The integration over the momentum must be replaced by the summation over the discrete momentum eigenvalues.
Since the eigenvalues are more separated as they increase, the resulting summation may converge even if the original integral diverges.
In Section \ref{eli}, we will demonstrate how typical, ultraviolet divergent integrals in Feynman diagrams of QED turn into convergent sums.

\section{Integration in noncommutative momentum space}
\label{nonco}

For general noncommutative algebras, one cannot separate the notations of trace and integral \cite{DoNe}.
The integration in a noncommutative space is associated with the summation over the operator eigenstates \cite{eigen}.
In order to find a physical prescription on how to integrate in the noncommutative momentum space, we consider the classical and quantum partition functions.

In classical statistical physics, the canonical-ensemble partition function is given by the integration of the exponent of the Hamiltonian $H(q_i,p_i)$ of the system over phase space:
\begin{equation}
Z=\frac{1}{h^s}\int dq_1\dots dq_s \int dp_1\dots dp_s\,  e^{-\frac{H(q,p)}{kT}},
\end{equation}
where $i=1,\dots,s$ counts the degrees of freedom, $q_i$ are the generalized coordinates, $p_i$ are the generalized momenta, $k$ is the Boltzmann constant, and $T$ is the temperature of the system represented by the canonical ensemble \cite{Patr}.
In quantum statistical physics, the partition function is the sum of the exponential of the energies of the eigenstates:
\begin{equation}
Z=\sum_i e^{-\frac{E_i}{kT}}.
\end{equation}
For example, for a one-dimensional harmonic oscillator with mass $m$ and angular frequency $\omega$, $H=p^2/(2m)+(1/2)m\omega^2 q^2$, and the classical partition function (the integration over $q$ and $p$ is from $-\infty$ to $\infty$) is $(kT)/(\hbar\omega)$.
The corresponding quantum partition function, using the energy eigenvalues of the harmonic oscillator $E_i=\hbar\omega(i+1/2)$ with integer $i$ running from 0 to $\infty$, is $(1/2)/\sinh[(\hbar\omega)/(kT)]$.
In the limit $\hbar\rightarrow 0$, the quantum partition function tends to its classical value.

The correspondence between the two partition functions is
\[
\frac{1}{2\pi\hbar}\int dq \int dp \leftrightarrow \sum_{\textrm{eigenstates}},
\]
or
\begin{equation}
\iint dq\,dp\, f(H(q,p)) \leftrightarrow 2\pi\sum_{\textrm{eigenstates}} f(E)\,|[q,p]|.
\label{partition}
\end{equation}
To account for the quantum commutation relation between the integration variables $q$ and $p$, which leads to a discrete spectrum of energy eigenstates, the integration over continuous phase space in the classical partition function must be replaced with the summation over the eigenstates.
The quantity under the sum is multiplied by the absolute value of the commutator of the integration variables, $|[q,p]|=\hbar$, and gives a more accurate, quantum partition function.

We propose that for three spatial components $n_i$ of a vector quantity ${\bf n}$, satisfying the cyclic commutation relations:
\begin{equation}
[n_x,n_y]=in_z,\,\,\,[n_y,n_z]=in_x,\,\,\,[n_z,n_x]=in_y,
\label{commut}
\end{equation}
the integration over the ${\bf n}$ space must be replaced with a summation over eigenstates:
\begin{equation}
\int dn_x \int dn_y \int dn_z\, f({\bf n}^2)\rightarrow 4\pi \sum_{\textrm{eigenstates}} f({\bf n}^2)\,|n_z|,
\label{pres}
\end{equation}
where $f({\bf n}^2)$ is any scalar function of the square of the vector, ${\bf n}^2$.
This prescription is analogous to the partition function (\ref{partition}): the integration over continuous space of $n_x$ and $n_y$ is replaced with the summation over the eigenstates, the quantity under the sum is multiplied by the absolute value of the commutator of the integration variables, $|[n_x,n_y]|=|n_z|$, and the integration over $n_z$ is also replaced with the summation over the eigenstates.

The summation is over the eigenvalues of ${\bf n}$, which are given by $|{\bf n}|=\sqrt{l(l+1)}$ and $n_z=m$, where $l$ is the orbital quantum number (a nonnegative integer) and $m$ is the magnetic quantum number (an integer $m\in[-l,l]$).
These quantum numbers follow from the quantum commutation relations (\ref{commut}) that are analogous to the commutation relations for the angular momentum.
We obtain for the prescription (\ref{pres}):
\begin{equation}
\iiint dn_x dn_y dn_z\, f({\bf n}^2)\rightarrow 4\pi\sum_{l=1}^\infty\sum_{m=-l}^l f({\bf n}^2)\,|m|=4\pi \sum_{l=1}^\infty f({\bf n}^2)\,l(l+1).
\label{prescription}
\end{equation}
The summation over $l$ is from 1 to $\infty$ since for $l=0$ we have $m=0$ and thus $n_z=0$, which does not contribute to (\ref{pres}).

In the noncommutative momentum space, we have the commutation relations (\ref{com}).
Without loss of generality, we assume $Q>0$ (the case $Q<0$ for the right-handed coordinate system is equivalent to the case $Q>0$ for the left-handed one).
We introduce a vector
\begin{equation}
{\bf n}=\frac{{\bf p}}{Q}
\label{vector}
\end{equation}
and denote
\[
n=|{\bf n}|.
\]
This vector satisfies the commutation relations (\ref{commut}).
In order to use the prescription (\ref{prescription}), we replace the integration over the momentum space with the integration over the ${\bf n}$ space:
\begin{equation}
\iiint dp_x dp_y dp_z\, f({\bf p}^2)\rightarrow \iiint dn_x dn_y dn_z J\,f(Q^2 n^2)\rightarrow 4\pi \sum_{l=1}^\infty J\,f(Q^2 n^2)\,l(l+1),
\label{prescr}
\end{equation}
where $J=\partial(p_x,p_y,p_z)/\partial(n_x,n_y,n_z)$ is the Jacobian of the transformation from the components of ${\bf p}$ to the components of ${\bf n}$ and $f({\bf p}^2)$ is any scalar function of ${\bf p}^2$.

As an example illustrating this correspondence, we consider a spatial Gaussian function $f({\bf p}^2)=e^{-k{\bf p}^2}$, where $k>0$ is a constant.
This function, integrated over the three components of the momentum, gives $(\pi/k)^{3/2}$.
If $Q$ is constant then $J=Q^3$ and the integral $\int e^{-k{\bf p}^2} d{\bf p}$ turns into
\[
\int e^{-kQ^2 n^2}Q^3 d{\bf n}\rightarrow 4\pi \sum_{l=1}^\infty e^{-kQ^2 l(l+1)}Q^3 l(l+1),
\]
where we denote $d{\bf p}=dp_x dp_y dp_z$ and $d{\bf n}=dn_x dn_y dn_z$.
This sum in the limit of continuous momentum space, $Q\rightarrow 0$, tends to $(\pi/k)^{3/2}$, which can be verified numerically or using the following limit:
\begin{equation}
\lim_{Q\to 0} \sum_{l=1}^\infty e^{-kQ^2 l(l+1)}Q^3 l(l+1)=\int_{\sqrt{2}}^\infty e^{-kQ^2 \eta^2}Q^3 \eta^2 d\eta=\int_0^\infty e^{-k\zeta^2}\zeta^2 d\zeta,
\label{limit}
\end{equation}
where $\zeta=Q\eta$.
In this limit, the separation between adjacent values of $l$ does not affect significantly the function $f$ and thus the summation over $l$ can be replaced with the integration over $\eta$ with $\eta=\sqrt{l(l+1)}$.

The quantity $Q$, however, depends on the momentum, following the dimensionality of the Cartan equations.
In this example, we take $Q=U|{\bf p}|^3$, where $U>0$ is a constant, which is the spatial case of (\ref{val}).
Accordingly, $Q=Un^3 Q^3$ and thus $Q=U^{-1/2}n^{-3/2}$.
Using $n^2=n_x^2+n_y^2+n_z^2$, we find $\partial p_x/\partial n_x=\partial(Qn_x)/\partial n_x=U^{-1/2}n^{-7/2}(n^2-3n_x^2/2)$, $\partial p_x/\partial n_y=\partial(Qn_x)/\partial n_y=U^{-1/2}n^{-7/2}(-3n_x n_y/2)$, and similarly for other partial derivatives.
The resulting Jacobian is $J=U^{-3/2}n^{-9/2}/2$ and the integral considered turns into
\begin{equation}
\int e^{-kQ^2 n^2}J\,d{\bf n}\rightarrow 2\pi U^{-3/2}\sum_{l=1}^\infty e^{-k/(Un)}n^{-9/2}l(l+1)=2\pi U^{-3/2}\sum_{l=1}^\infty e^{-k/[U\sqrt{l(l+1)}]}[l(l+1)]^{-5/4}. \nonumber
\end{equation}
This sum in the limit of continuous momentum space, $U\rightarrow 0$, also tends to $(\pi/k)^{3/2}$, which can be verified numerically or using the limit (\ref{limit}):
\begin{equation}
\lim_{U\to 0} U^{-3/2}\sum_{l=1}^\infty e^{-k/[U\sqrt{l(l+1)}]}[l(l+1)]^{-5/4}=U^{-3/2}\int_{\sqrt{2}}^\infty e^{-k/(U\eta)}\eta^{-5/2}d\eta=2\int_0^\infty e^{-k\zeta^2}\zeta^2 d\zeta,
\end{equation}
where $\zeta=1/\sqrt{U\eta}$.

\section{Elimination of ultraviolet divergences}
\label{eli}

A typical, logarithmically divergent integral in QED has a form $(2\pi)^{-4}\int d^4p/(p^2-\mu^2+i\epsilon)^2$, where $\mu$ does not depend on $p$.
If we assume that the four-momentum space is not curved, then the integration is taken in the four-momentum space with the Lorentzian (pseudo-Euclidean) metric.
Hereinafter, we use units such that $\hbar=1$.
Calculations are simplified by the Wick rotation: the time component of the momentum $p^0$ is replaced with $ip^0_\textrm{E}$ \cite{qft}.
Accordingly, the integration is taken in the Euclidean four-momentum space, with $p^2$ replaced with $p^2_\textrm{E}=(p^0)^2_\textrm{E}+{\bf p}^2_\textrm{E}$.
The integral becomes $i(2\pi)^{-4}\int d^4p_\textrm{E}/(p^2_\textrm{E}+\mu^2)^2$.\footnote{
We assume that the Feynman propagator for a Dirac particle with four-momentum $p_\mu$ and mass $m$ is $\tilde{S}_\textrm{F}(p)=i(\gamma^\mu p_\mu+m)/(p^2-m^2+i\epsilon)$ and for a photon (in the Feynman gauge) is $\tilde{D}^{\mu\nu}_\textrm{F}(p)=-ig^{\mu\nu}/(p^2+i\epsilon)$, where $\gamma^\mu$ are the Dirac matrices in flat spacetime (with $g_{\mu\nu}$ being the Minkowski metric tensor): $\gamma^{(\mu}\gamma^{\nu)}=g^{\mu\nu}I_4$, $p^2=p_\mu p^\nu$, and $\epsilon\to 0^{+}$ \cite{qft}.
The Fourier transform of a propagator in the four-momentum representation, with the exponential factor $e^{-ip_\mu(x-y)^\mu}$, gives the position representation of the propagator ($S_\textrm{F}(x-y)$ or $D_\textrm{F}(x-y)$) describing a particle moving from one point $x$ in spacetime to another point $y$ that is infinitesimally close to $x$.
In the Fourier transform, the integration over the four-momentum must be replaced with the summation over the four-momentum eigenvalues, according to the presented prescription.
A propagator describing the motion between two points separated by a finite distance can be constructed as a sequence of infinitesimal propagators using the Riemann normal coordinates, which depend on the curvature and torsion tensors \cite{Par}.}

Applying the Pauli--Villars regularization to this integral (and omitting the factor $i$ and subscript E), one obtains \cite{PaVi}
\begin{eqnarray}
& & \frac{1}{(2\pi)^4}\Bigl[\int\frac{d^4p}{(p^2+\mu^2)^2}-\int\frac{d^4p}{(p^2+\Lambda^2)^2}\Bigr]=\frac{1}{(2\pi)^4}(2\pi^2)\Bigl[\int_0^\infty\frac{p^3dp}{(p^2+\mu^2)^2}-\int_0^\infty\frac{p^3dp}{(p^2+\Lambda^2)^2}\Bigr] \nonumber \\
& & =\frac{1}{16\pi^2}\Bigl[\int_0^\infty\frac{x\,dx}{(x+\mu^2)^2}-\int_0^\infty\frac{x\,dx}{(x+\Lambda^2)^2}\Bigr]=\frac{1}{16\pi^2}\Bigl[\ln\frac{x+\mu^2}{x+\Lambda^2}+\frac{\mu^2}{x+\mu^2}-\frac{\Lambda^2}{x+\Lambda^2}\Bigr]\Big|_0^\infty=-\frac{1}{16\pi^2}\ln\frac{\mu^2}{\Lambda^2},
\label{Pauli}
\end{eqnarray}
since the volume of the hypersurface of a four-sphere of radius $p$ is $2\pi^2 p^3$.
Applying the dimensional regularization to this integral gives \cite{HoVe}
\[
\frac{1}{(2\pi)^n}\int\frac{d^np}{(p^2+\mu^2)^2}=\frac{1}{8\pi^2\epsilon}-\frac{1}{16\pi^2}\Bigl(\ln\frac{\mu^2}{4\pi}+\gamma\Bigr)+O(\epsilon),
\]
where $\gamma$ is the Euler--Mascheroni constant.
This result, in the limit $\epsilon\rightarrow 0$, is consistent with that in the Pauli--Villars regularization if $1/\epsilon+[\ln(4\pi)-\gamma]/2$ is identified with $\ln\Lambda$.

We now apply the torsional regularization to this integral, using the prescription (\ref{prescr}).
Introducing a vector ${\bf n}={\bf p}/Q$ as in (\ref{vector}) and the Jacobian $J=\partial(p_x,p_y,p_z)/\partial(n_x,n_y,n_z)$, we obtain
\begin{equation}
\int\frac{d^4p}{(p^2+\mu^2)^2}=\int\frac{dp_0\,d{\bf p}}{(p^2+\mu^2)^2}=\int\frac{dp_0\,J\,d{\bf n}}{(p^2+\mu^2)^2}\rightarrow 4\pi\int_{-\infty}^\infty dp_0 \sum_{l=1}^\infty\frac{J}{(p^2+\mu^2)^2}\,l(l+1),
\label{integ}
\end{equation}
where $p^2=p_0^2+Q^2 n^2=p_0^2+Q^2 l(l+1)$.
If $Q$ is constant then $J=Q^3$ and the sum-integral in (\ref{integ}) is
\[
4\pi Q^3\sum_{l=1}^\infty\int_{-\infty}^\infty dp_0\frac{l(l+1)}{[p_0^2+Q^2 l(l+1)+\mu^2]^2}=2\pi^2 Q^3 \sum_{l=1}^\infty\frac{l(l+1)}{[Q^2 l(l+1)+\mu^2]^{3/2}}=2\pi^2 \sum_{l=1}^\infty\frac{l(l+1)}{[l(l+1)+\mu^2/Q^2]^{3/2}},
\nonumber
\]
which diverges as $\sim\sum_{l=1}^\infty l^{-1}$.

However, the dimensionality of the Cartan equations implies (\ref{val}).
Without loss of generality, we assume $U>0$.
We therefore have
\begin{equation}
p^2=p_0^2+U^2 n^2 p^6.
\label{condit}
\end{equation}
In order to find the Jacobian $J$ of the transformation from ${\bf p}$ to ${\bf n}$, we proceed as follows.
Differentiating (\ref{condit}) with respect to $n_x$ gives $2p\frac{\partial p}{\partial n_x}=6U^2 n^2 p^5 \frac{\partial p}{\partial n_x}+2U^2 p^6 n_x$ and thus
\[
\frac{\partial p}{\partial n_x}=\frac{U^2 p^5 n_x}{1-3U^2 n^2 p^4}.
\]
Consequently, we find
\begin{eqnarray}
& & \frac{\partial p_x}{\partial n_x}=\frac{\partial(Qn_x)}{\partial n_x}=Q+3Un_x p^2 \frac{\partial p}{\partial n_x}=\frac{Q}{1-3U^2 n^2 p^4}[1-3U^2 p^4 (n_y^2 + n_z^2)], \nonumber \\
& & \frac{\partial p_x}{\partial n_y}=\frac{\partial(Qn_x)}{\partial n_y}=3Un_x p^2 \frac{\partial p}{\partial n_y}=\frac{Q}{1-3U^2 n^2 p^4}(3U^2 p^4 n_x n_y), \nonumber
\end{eqnarray}
and similarly for other components.
The Jacobian is
\[
J=\mbox{det}\left( \begin{array}{ccc}
\partial p_x/\partial n_x & \partial p_x/\partial n_y & \partial p_x/\partial n_z \\
\partial p_y/\partial n_x & \partial p_y/\partial n_y & \partial p_y/\partial n_z \\
\partial p_z/\partial n_x & \partial p_z/\partial n_y & \partial p_z/\partial n_z \end{array} \right)=\frac{Q^3}{1-3U^2 n^2 p^4}.
\]
Substituting $J$ into the sum-integral in (\ref{integ}), using $dp_0/dp=(1-3U^2 n^2 p^4)/(1-U^2 n^2 p^4)^{1/2}$, which results from (\ref{condit}), and taking the eigenvalues $n=\sqrt{l(l+1)}$ gives
\begin{eqnarray}
& & 4\pi\int_{-\infty}^\infty dp_0 \sum_{l=1}^\infty\frac{Q^3 n^2}{(1-3U^2 n^2 p^4)(p^2+\mu^2)^2}=4\pi\int dp \frac{dp_0}{dp} \sum_{l=1}^\infty\frac{Q^3 n^2}{(1-3U^2 n^2 p^4)(p^2+\mu^2)^2} \nonumber \\
& & =4\pi\int_{-1/\sqrt{Un}}^{1/\sqrt{Un}} dp \sum_{l=1}^\infty\frac{Q^3 n^2}{(1-U^2 n^2 p^4)^{1/2}(p^2+\mu^2)^2}=8\pi\int_0^{1/\sqrt{Un}} dp \sum_{l=1}^\infty\frac{U^3 p^9 n^2}{(1-U^2 n^2 p^4)^{1/2}(p^2+\mu^2)^2} \nonumber \\
& & =8\pi\int_0^1 d\xi \sum_{l=1}^\infty\frac{U^3 \xi^9 n^2 (Un)^{-5}}{(1-\xi^4)^{1/2}[\xi^2/(Un)+\mu^2]^2}=8\pi\int_0^1 d\xi \sum_{l=1}^\infty\frac{\xi^9 n^{-1}}{(1-\xi^4)^{1/2}[\xi^2+U\mu^2 n]^2} \nonumber \\
& & =4\pi\int_0^1 d\zeta \sum_{l=1}^\infty\frac{\zeta^4 n^{-1}}{(1-\zeta^2)^{1/2}[\zeta+U\mu^2 n]^2}=4\pi\sum_{l=1}^\infty \int_0^{\pi/2} d\phi \frac{\sin^4\phi\,n^{-1}}{[\sin\phi+U\mu^2 n]^2} \nonumber \\
& & =4\pi\sum_{l=1}^\infty \int_0^{\pi/2} d\phi \frac{\sin^4\phi\,[l(l+1)]^{-1/2}}{[\sin\phi+U\mu^2\sqrt{l(l+1)}]^2},
\label{sumint}
\end{eqnarray}
where we denote $Unp^2=\xi^2=\zeta=\sin\phi$.

The sum-integral (\ref{sumint}) converges, which follows from its behavior at large values of $l$: $\sim\sum_{l=1}^\infty l^{-3}$.
It depends only on the nondimensional quantity $U\mu^2$.
Its value, divided by $(2\pi)^4$, gives the torsional-regularized, finite value of the integral $(2\pi)^{-4}\int d^4p_\textrm{E}/(p^2_\textrm{E}+\mu^2)^2$ in the noncommutative momentum space resulting from torsion.
We note that regularization is possible because $Q$ appearing in the commutation relations for the momentum (\ref{com}) is not constant but increases with the magnitude of the four-momentum.
Consequently, torsional regularization works for a typical, logarithmically divergent integral in QED.

Now, we consider the limit of continuous four-momentum space for the sum-integral (\ref{sumint}) using (\ref{limit}).
If $U=0$ (no torsion), this sum-integral diverges as $\sim\sum_{l=1}^\infty l^{-1}$, as expected.
In the limit $U\rightarrow 0$, this sum-integral diverges as $\ln{U}$:
\begin{eqnarray}
& & 4\pi\int_{\sqrt{2}}^\infty d\eta \int_0^{\pi/2} d\phi \frac{\sin^4\phi\,\eta^{-1}}{[\sin\phi+U\mu^2\eta]^2}=4\pi\int_0^{\pi/2} d\phi\,\sin^4\phi \int_{{\sqrt{2}}U\mu^2}^\infty d\zeta \frac{\zeta^{-1}}{[\sin\phi+\zeta]^2} \nonumber \\
& & =4\pi\int_0^{\pi/2} d\phi\,\sin^2\phi \Bigl[\ln\frac{\zeta}{\zeta+\sin\phi}+\frac{\sin\phi}{\zeta+\sin\phi}\Bigr]\Big|_{\zeta={\sqrt{2}}U\mu^2}^{\zeta\to\infty}=-4\pi\int_0^{\pi/2} d\phi\,\sin^2\phi [\ln(U\mu^2)+\ln{\sqrt{2}}-\ln\sin\phi+1] \nonumber \\
& & =-\pi^2\ln(U\mu^2)+\mbox{finite terms}, \nonumber
\end{eqnarray}
where we denote $\zeta=U\mu^2\eta$.
Its divergent part, divided by $(2\pi)^4$, is equal to $-1/(16\pi^2)\ln(U\mu^2)$, which is the same as the value in (\ref{Pauli}) if $U=1/\Lambda^2$.
Accordingly, $\Lambda^2$ in Pauli--Villars regularization can be identified as an effective $1/U$.\footnote{
Our regularization prescription does not introduce an explicit momentum space cutoff $\Lambda$ and thus does not have ultraviolet-infrared divergence mixing that occurs for field theories with noncommutative coordinate space that do have such a cutoff \cite{DoNe}.
}
Consequently, the fictitious mass $\Lambda$ is on the order of the Planck mass.
This order of magnitude originates from the spin-torsion coupling in EC.
Pauli--Villars regularization can thus be regarded as a mathematical technique that is equivalent to the realistic, torsional regularization.

We now generalize the prescription (\ref{sumint}) to an integral of form $(2\pi)^{-4}\int d^4p/(p^2-\mu^2+i\epsilon)^s$, where $s$ is an arbitrary positive integer.
If $s=1$ or $s=2$ (which is the case considered previously), such an integral is divergent.
If $s\ge 3$, such an integral has a finite value.
In the Euclidean four-momentum space, this integral becomes $(-1)^s i(2\pi)^{-4}\int d^4p_\textrm{E}/(p^2_\textrm{E}+\mu^2)^s$.
Following the steps leading to (\ref{sumint}), we obtain
\begin{eqnarray}
& & \int\frac{d^4p}{(p^2+\mu^2)^s}\rightarrow 8\pi\int_0^{1/\sqrt{Un}} dp \sum_{l=1}^\infty\frac{U^3 p^9 n^2}{(1-U^2 n^2 p^4)^{1/2}(p^2+\mu^2)^s} \nonumber \\
& & =8\pi\int_0^1 d\xi \sum_{l=1}^\infty\frac{U^3 \xi^9 n^2 (Un)^{-5}}{(1-\xi^4)^{1/2}[\xi^2/(Un)+\mu^2]^s}=8\pi\int_0^1 d\xi \sum_{l=1}^\infty\frac{U^{s-2}\xi^9 n^{s-3}}{(1-\xi^4)^{1/2}[\xi^2+U\mu^2 n]^s} \nonumber \\
& & =4\pi\int_0^1 d\zeta \sum_{l=1}^\infty\frac{U^{s-2}\zeta^4 n^{s-3}}{(1-\zeta^2)^{1/2}[\zeta+U\mu^2 n]^s}=4\pi U^{s-2}\sum_{l=1}^\infty \int_0^{\pi/2} d\phi \frac{\sin^4\phi\,n^{s-3}}{[\sin\phi+U\mu^2 n]^s} \nonumber \\
& & =4\pi U^{s-2}\sum_{l=1}^\infty \int_0^{\pi/2} d\phi \frac{\sin^4\phi\,[l(l+1)]^{(s-3)/2}}{[\sin\phi+U\mu^2\sqrt{l(l+1)}]^s}.
\label{sumintsc}
\end{eqnarray}

The sum-integral (\ref{sumintsc}) converges, which follows from its behavior at large values of $l$: $\sim\sum_{l=1}^\infty l^{-3}$.
Its value, divided by $(2\pi)^4$, gives the torsional-regularized, finite value of the integral $(2\pi)^{-4}\int d^4p_\textrm{E}/(p^2_\textrm{E}+\mu^2)^s$ in the noncommutative momentum space resulting from torsion.
Consequently, torsional regularization works not only for logarithmically divergent integrals ($s=2$), but also for quadratically divergent integrals ($s=1$) that appear in QED.

To verify that the proposed regularization procedure reproduces the values of finite integrals in the limit of continuous four-momentum space ($U\to 0$), we apply it to the integral in (\ref{sumintsc}) for $s=3$, whose value is $\pi^2/(2\mu^2)$.
We obtain
\begin{eqnarray}
& & \int\frac{d^4p}{(p^2+\mu^2)^3}\rightarrow 4\pi U\sum_{l=1}^\infty \int_0^{\pi/2} d\phi \frac{\sin^4\phi}{[\sin\phi+U\mu^2\sqrt{l(l+1)}]^3} \nonumber \\
& & \to 4\pi U\int_0^{\pi/2} d\phi \int_{\sqrt{2}}^\infty d\eta \frac{\sin^4\phi}{[\sin\phi+U\mu^2\eta]^3}=\frac{4\pi}{\mu^2}\int_0^{\pi/2} d\phi \int_0^\infty d\zeta \frac{\sin^4\phi}{[\sin\phi+\zeta]^3}=\frac{4\pi}{\mu^2}\int_0^{\pi/2} d\phi\frac{\sin^2\phi}{2}=\frac{\pi^2}{2\mu^2}. \nonumber
\end{eqnarray}

\section{Tensor integrals}

The integrals considered in the preceding section are scalar integrals.
We now consider a tensor integral $\int d^4p_\textrm{E}\,p^\mu_\textrm{E} p^\nu_\textrm{E}/(p^2_\textrm{E}+\Delta)^s$, where $\Delta>0$ does not depend on $p$ (and is equivalent to $\mu^2$ is the preceding section).
Such an integral appears in the Feynman diagram representing vacuum polarization \cite{qft}.
We omit the subscript E and use the identity
\begin{equation}
\int d^4p\frac{\partial}{\partial p_\nu}\Bigl(\frac{p^\mu}{(p^2+\Delta)^s}\Bigr)=\int d^4p\frac{\delta^{\mu\nu}}{(p^2+\Delta)^s}-2s\int d^4p\frac{p^\mu p^\nu}{(p^2+\Delta)^{s+1}},
\nonumber
\end{equation}
where $\delta^{\mu\nu}$ is the metric tensor in the four-dimensional Euclidean space.
The left-hand side of this equation can be transformed, Using the four-dimensional Gauss theorem, into a hypersurface integral that vanishes if the integration over $d^4p$ has no boundaries (as in Feynman diagrams) and if the integrals on the right-hand side have finite values \cite{vac}.
Consequently, this hypersurface integral should also vanish for regularized integrals.
We therefore have
\begin{equation}
\int d^4p\frac{p^\mu p^\nu}{(p^2+\Delta)^s}=\frac{\delta^{\mu\nu}}{2(s-1)}\int d^4p\frac{1}{(p^2+\Delta)^{s-1}}.
\label{iden}
\end{equation}

Using (\ref{sumintsc}) and (\ref{iden}), we obtain the torsional-regularized, finite values of typical integrals appearing in Feynman diagrams, in the noncommutative momentum space resulting from torsion:
\begin{eqnarray}
& & \int\frac{d^4p}{(p^2+\Delta)^s}\rightarrow 4\pi U^{s-2}\sum_{l=1}^\infty \int_0^{\pi/2} d\phi \frac{\sin^4\phi\,n^{s-3}}{[\sin\phi+U\Delta n]^s},\quad\int\frac{d^4p\,p^\mu}{(p^2+\Delta)^s}\rightarrow 0, \nonumber \\
& & \int\frac{d^4p\,p^\mu p^\nu}{(p^2+\Delta)^s}\rightarrow \frac{2\pi U^{s-3}\delta^{\mu\nu}}{s-1}\sum_{l=1}^\infty \int_0^{\pi/2} d\phi \frac{\sin^4\phi\,n^{s-4}}{[\sin\phi+U\Delta n]^{s-1}},
\label{sumintten}
\end{eqnarray}
with $n=\sqrt{l(l+1)}$.
The second integral in (\ref{sumintten}) is zero because of symmetry.
Accordingly, the following integral vanishes:
\begin{eqnarray}
& & \int d^4p\frac{-2p^\mu p^\nu+(p^2+\Delta)\delta^{\mu\nu}}{(p^2+\Delta)^2}=-2\int\frac{d^4p\,p^\mu p^\nu}{(p^2+\Delta)^2}+\delta^{\mu\nu}\int\frac{d^4p}{p^2+\Delta} \nonumber \\
& & \to -4\pi U^{-1}\delta^{\mu\nu}\sum_{l=1}^\infty \int_0^{\pi/2} d\phi \frac{\sin^4\phi\,n^{-2}}{\sin\phi+U\Delta n}+4\pi U^{-1}\delta^{\mu\nu}\sum_{l=1}^\infty \int_0^{\pi/2} d\phi \frac{\sin^4\phi\,n^{-2}}{\sin\phi+U\Delta n}=0,
\label{vanish}
\end{eqnarray}
in accordance with (\ref{iden}).
This integral also vanishes in dimensional regularization \cite{HoVe}.
Since the loop integrals at each order of perturbation theory in QED are of form (\ref{sumintten}) (with more vector indices), torsional regularization should give finite results at all orders.

\section{Vacuum polarization}

In this section, we apply torsional regularization to vacuum polarization (photon self-energy) in QED \cite{Schw}.
The aim of this calculation is a finite value of the bare charge of an electron.
The determination of this quantity should elucidate the nature of the fine-structure constant.

The vacuum polarization tensor at one-loop order for a photon with four-momentum $q$ creating an electron-positron bubble is given by \cite{qft}
\begin{eqnarray}
& & i\Pi^{\mu\nu}_\textrm{bubble}(q)=-e^2_0\int\frac{d^4k}{(2\pi)^4}\textrm{Tr}\Bigl[\gamma^\mu\frac{\gamma^\rho k_\rho+m}{k^2-m^2+i\epsilon}\gamma^\nu\frac{\gamma^\sigma(q+k)_\sigma+m}{(q+k)^2-m^2+i\epsilon}\Bigr] \nonumber \\
& & =-4e^2_0\int\frac{d^4k}{(2\pi)^4}\int_0^1 dx\frac{k^\mu(q+k)^\nu+k^\nu(q+k)^\mu-g^{\mu\nu}k\cdot(q+k)+m^2 g^{\mu\nu}}{(k^2+2q\cdot kx+q^2 x-m^2+i\epsilon)^2} \nonumber \\
& & =-\frac{\alpha_0}{\pi^3}\int d^4p \int_0^1 dx\frac{2p^\mu p^\nu-p^2 g^{\mu\nu}+\Delta g^{\mu\nu}+2(q^2 g^{\mu\nu}-q^\mu q^\nu)x(1-x)}{(p^2-\Delta+i\epsilon)^2}, \nonumber
\end{eqnarray}
where $e_0$ is the absolute value of the (bare) electric charge of an electron, $m$ is the (dressed) mass of an electron, $\alpha_0=e^2_0/(4\pi)$ is the (bare) fine-structure constant, $p=k+qx$, $x$ is the Feynman parameter, and
\[
\Delta=m^2-q^2 x(1-x).
\]
For $q^2<4m^2$, below the threshold for the production of an electron-positron pair, $\Delta>0$.
Applying the Wick rotation to this integral and using the identity (\ref{vanish}) gives
\begin{eqnarray}
& & \Pi^{\mu\nu}_\textrm{bubble}(q)=-\frac{\alpha_0}{\pi^3}\int d^4p_\textrm{E} \int_0^1 dx\frac{-2p^\mu_\textrm{E} p^\nu_\textrm{E}+p^2_\textrm{E} \delta^{\mu\nu}+\Delta \delta^{\mu\nu}+2(q^2 g^{\mu\nu}-q^\mu q^\nu)x(1-x)}{(p^2_\textrm{E}+\Delta)^2} \nonumber \\
& & =-\frac{2\alpha_0}{\pi^3}\int d^4p_\textrm{E} \int_0^1 dx\frac{x(1-x)}{(p^2_\textrm{E}+\Delta)^2}(q^2 g^{\mu\nu}-q^\mu q^\nu)=\Pi(q^2)q^2\Bigl(g^{\mu\nu}-\frac{q^\mu q^\nu}{q^2}\Bigr),
\label{trans}
\end{eqnarray}
where
\begin{equation}
\Pi(q^2)=-\frac{2\alpha_0}{\pi^3}\int d^4p_\textrm{E} \int_0^1 dx\frac{x(1-x)}{(p^2_\textrm{E}+\Delta)^2}.
\label{Pi}
\end{equation}
The tensor (\ref{trans}) is transverse: $\Pi^{\mu\nu}_\textrm{bubble}(q)q_\nu=0$, which follows from the Ward identity (conservation of charge).
This tensor is thus gauge invariant and the photon remains massless \cite{qft}.

We now apply torsional regularization to (\ref{Pi}).
Using the prescription (\ref{sumint}), or for the first integral in (\ref{sumintten}) for $s=2$, we obtain
\begin{equation}
\Pi(q^2)\rightarrow -\frac{8\alpha_0}{\pi^2}\sum_{l=1}^\infty \int_0^1 dx \int_0^{\pi/2} d\phi \frac{\sin^4\phi\,n^{-1}x(1-x)}{[\sin\phi+U\Delta n]^2}=-\frac{8\alpha_0}{\pi^2}\sum_{l=1}^\infty \int_0^1 dx \int_0^{\pi/2} d\phi \frac{\sin^4\phi\,[l(l+1)]^{-1/2}x(1-x)}{[\sin\phi+U\Delta\sqrt{l(l+1)}]^2}.
\label{Pito}
\end{equation}
Consequently, $\Pi(q^2)$ has a finite value (if $\alpha_0$ is finite) since the sum in (\ref{Pito}) converges as $\sim\sum_{l=1}^\infty l^{-3}$.
Vacuum polarization in QED at one-loop order, with torsional regularization, therefore gives a finite, gauge-invariant correction to the transverse part of the photon propagator.

The sum-integral (\ref{Pito}) can be rewritten as
\[
\Pi(q^2)=\Pi(0)+\delta\Pi(q^2),
\]
where
\begin{equation}
\Pi(0)=-\frac{8\alpha_0}{\pi^2}\sum_{l=1}^\infty \int_0^1 dx \int_0^{\pi/2} d\phi \frac{\sin^4\phi\,n^{-1}x(1-x)}{[\sin\phi+Um^2 n]^2}.
\label{Pi0}
\end{equation}
Summing over the one-particle-irreducible contributions $\Pi$ to the photon propagator gives $1+\Pi+\Pi^2+\dots=1/(1-\Pi)$.
Since $\delta\Pi(q^2)$ is proportional to $\alpha_0$, $\Pi(0)$ can be absorbed into a renormalized (dressed) value of the fine-structure constant $\alpha$ (on-shell renormalization) \cite{qft,Dys,Gel}:
\[
\alpha_\textrm{run}=\frac{\alpha_0}{1-\Pi(q^2)}=\frac{\alpha_0}{(1-\Pi(0))\bigl(1-\frac{\delta\Pi(q^2)}{1-\Pi(0)}\bigr)}=\frac{\alpha}{1-\delta\Pi_\textrm{R}(q^2)},
\]
where
\begin{equation}
\alpha=\frac{\alpha_0}{1-\Pi(0)},\quad\Pi_\textrm{R}(q^2)=\frac{\alpha}{\alpha_0}\Pi(q^2),\quad\delta\Pi_\textrm{R}(q^2)=\frac{\alpha}{\alpha_0}\delta\Pi(q^2).
\label{reno}
\end{equation}
Subtracting $\Pi(0)$ from $\Pi(q^2)$ is turned into dividing $\alpha_0$ by $1-\Pi(0)$ \cite{Hua}, which was proved by Dyson to all orders of perturbation theory \cite{Dys}.
The running coupling in QED is therefore
\begin{equation}
\alpha_\textrm{run}=\alpha\Bigl[1+\frac{8\alpha}{\pi^2}\sum_{l=1}^\infty \int_0^1 dx \int_0^{\pi/2} d\phi \frac{\sin^4\phi\,n^{-1}x(1-x)}{[\sin\phi+U\Delta n]^2}-\frac{8\alpha}{\pi^2}\sum_{l=1}^\infty \int_0^1 dx \int_0^{\pi/2} d\phi \frac{\sin^4\phi\,n^{-1}x(1-x)}{[\sin\phi+Um^2 n]^2}\Bigr]^{-1},
\label{run}
\end{equation}
where $\alpha=\alpha_\textrm{run}(q^2=0)\approx 1/137.036$ \cite{PDG}.

For a virtual photon in scattering of two charged particles, $q^2<0$.
In the static limit, giving the one-loop quantum correction to the Coulomb potential, $q^0=0$ and $q^2=-{\bf q}^2$, and thus $\Delta=m^2+{\bf q}^2 x(1-x)$.
As ${\bf q}^2$ increases, so does $\alpha_\textrm{run}$ in (\ref{run}).
There is no Landau pole if $\alpha_\textrm{run}$ stays finite as ${\bf q}^2$ increases to infinity.
This condition is guaranteed if
\begin{equation}
-\Pi_\textrm{R}(0)=\frac{8\alpha}{\pi^2}\sum_{l=1}^\infty \int_0^1 dx \int_0^{\pi/2} d\phi \frac{\sin^4\phi\,n^{-1}x(1-x)}{[\sin\phi+Um^2 n]^2}<1.
\label{Land}
\end{equation}

The relations (\ref{Pi0}) and (\ref{reno}) give
\begin{equation}
\alpha_0=\frac{\alpha}{1+\Pi_\textrm{R}(0)}=\alpha\Bigl[1-\frac{8\alpha}{\pi^2}\sum_{l=1}^\infty \int_0^1 dx \int_0^{\pi/2} d\phi \frac{\sin^4\phi\,n^{-1}x(1-x)}{[\sin\phi+Um^2 n]^2}\Bigr]^{-1}.
\nonumber
\end{equation}
Equivalently, the absolute values of the bare charge $e_0$ and the renormalized (observed) charge $e=\sqrt{4\pi\alpha}$ are related by
\begin{equation}
e_0=\frac{e}{(1+\Pi_\textrm{R}(0))^{1/2}}=e\Bigl[1-\frac{8\alpha}{\pi^2}\sum_{l=1}^\infty \int_0^1 dx \int_0^{\pi/2} d\phi \frac{\sin^4\phi\,n^{-1}x(1-x)}{[\sin\phi+Um^2 n]^2}\Bigr]^{-1/2}.
\label{bare}
\end{equation}
The bare charge is finite if the condition (\ref{Land}) is satisfied.
So is the renormalization constant for the photon wave function \cite{qft}:
\[
Z_3=\frac{e^2}{e^2_0}.
\]

In the low-energy limit (relative to the Planck energy), when $|q^2|\ll U^{-2}$, we can approximate in (\ref{Pito}) the summation over $l$ with the integration over $\eta=\sqrt{l(l+1)}$, as in (\ref{limit}):
\begin{eqnarray}
& & \Pi(q^2)\approx -\frac{8\alpha_0}{\pi^2} \int_0^1 dx \int_0^{\pi/2} d\phi \int_{\sqrt{2}}^\infty d\eta \frac{\sin^4\phi\,\eta^{-1}x(1-x)}{[\sin\phi+U\Delta\eta]^2}=-\frac{8\alpha_0}{\pi^2} \int_0^1 dx \int_0^{\pi/2} d\phi \int_{\sqrt{2}U\Delta}^\infty d\zeta \frac{\sin^4\phi\,\zeta^{-1}x(1-x)}{[\sin\phi+\zeta]^2} \nonumber \\
& & \approx -\frac{8\alpha_0}{\pi^2} \int_0^1 dx\,x(1-x) \int_0^{\pi/2} d\phi\,\sin^2\phi\Bigl(\ln\frac{\sin\phi}{\sqrt{2}U\Delta}-1\Bigr) \nonumber \\
& & =-\frac{8\alpha_0}{\pi^2} \int_0^1 dx\,x(1-x)\Bigl(N-\frac{\pi}{4}\ln(Um^2)-\frac{\pi}{4}(\ln\sqrt{2}+1)-\frac{\pi}{4}\ln\frac{\Delta}{m^2}\Bigr) \nonumber \\
& & =\frac{2\alpha_0}{\pi}\int_0^1 dx\,x(1-x)\ln\Bigl(1-\frac{q^2}{m^2}x(1-x)\Bigr)+\frac{\alpha_0}{3\pi}\Bigl(\ln(Um^2)+\ln\sqrt{2}+1-\frac{4N}{\pi}\Bigr),
\end{eqnarray}
where $N=\int_0^{\pi/2}\sin^2\phi\,\ln\sin\phi\,d\phi\approx -0.151697$.
Consequently,
\begin{eqnarray}
& & \Pi_\textrm{R}(0)\approx\frac{\alpha}{3\pi}\Bigl(\ln(Um^2)+\ln\sqrt{2}+1-\frac{4N}{\pi}\Bigr), \nonumber \\
& & \delta\Pi_\textrm{R}(q^2)\approx\frac{2\alpha}{\pi}\int_0^1 dx\,x(1-x)\ln\Bigl(1-\frac{q^2}{m^2}x(1-x)\Bigr).
\label{appr}
\end{eqnarray}
The first formula in (\ref{appr}) is valid when $m^2\ll U^{-2}$, which holds for all known charged fermions.
The running coupling (\ref{run}) for $q^2=-{\bf q}^2$ is therefore approximated by the known formula \cite{qft}:
\[
\alpha_\textrm{run}\approx\alpha\Bigl[1-\frac{2\alpha}{\pi}\int_0^1 dx\,x(1-x)\ln\Bigl(1+\frac{{\bf q}^2}{m^2}x(1-x)\Bigr)\Bigr]^{-1},
\]
giving the Uehling potential that modifies the Coulomb potential $1/(4\pi r)$ at the one-loop order \cite{Ueh}:
\[
V(r)=\int\frac{d{\bf q}}{(2\pi)^3}e^{i{\bf q}\cdot{\bf r}}\frac{1}{{\bf q}^2(1-\delta\Pi_\textrm{R}(-{\bf q^2}))}.
\]

Finally, we determine the value of the bare charge of an electron.
We take $U=m^{-2}_\textrm{P}$.
If an electron were the only charged fermion, the first formula in (\ref{appr}) would give $\Pi_\textrm{R}(0)\approx -0.0786$, the condition (\ref{Land}) for the absence of a Landau pole would be satisfied and (\ref{bare}) would give $e_0\approx 1.042\,e$.
Taking the reduced Planck mass $m_\textrm{P}/\sqrt{8\pi}$ instead of the Planck mass in $U$ would give $\Pi_\textrm{R}(0)\approx -0.0761$ and $e_0\approx 1.040\,e$.
We note that the exact value of $U$, as long as it is on the order of the inverse squared Planck mass, is not too significant.
Including all charged fermions (denoted f) contributing to vacuum polarization replaces $\Pi_\textrm{R}(0)$ in (\ref{appr}) with
\[
\Pi_\textrm{R}(0)\approx\sum_\textrm{f}\frac{\alpha}{3\pi}\Bigl(\ln(Um^2_\textrm{f})+\ln\sqrt{2}+1-\frac{4N}{\pi}\Bigr)k_\textrm{f},
\]
where $k_\textrm{f}$ is the squared electric charge of a fermion (in the units of the charge of an electron).
Substituting the masses and charges of quarks and charged leptons \cite{PDG} gives $\Pi_\textrm{R}(0)\approx -0.328$, the condition (\ref{Land}) for the absence of a Landau pole is satisfied, and (\ref{bare}) gives
\[
e_0\approx 1.220\,e.
\]
The corresponding renormalization constant for the photon wave function is $Z_3=0.672$, which is independent of $\alpha$, in agreement with the behavior found in \cite{Gel}.
The corresponding bare fine-structure constant is $\alpha_0\approx 1/92.1$.
These results show that the bare charge of an electron (and thus $\alpha_0$) and the renormalization constant for the photon wave function are finite.
Consequently, including spin-torsion coupling naturally provides a physical regularization of ultraviolet divergences and makes the renormalization procedure finite: calculations involving the bare charge have a physical meaning.
Accordingly, QED with torsion could be ultraviolet complete.

The factor $1.220$ is universal for all fermions: the bare charge of each fermion is approximately $1.220$ times its effective (renormalized) charge.
The finite screening of an electric charge by virtual particle-antiparticle pairs does not depend on its value.
For example, an electron has bare charge $-1.22\,e$, a positron has bare charge $1.22\,e$, and a quark $u$ has bare charge $1.22\,(2/3)e$.
The bare charge of an electron is thus a constant of Nature and its exact determination should elucidate the origin of the fine-structure constant.

\section{Summary}

In this article, we showed that the noncommutativity of the four-dimensional momentum, resulting from torsion coupled to spin, regularizes a typical, ultraviolet divergent integral that appears in loop corrections to Feynman diagrams in QED.
In the noncommutative momentum space, the integration over the momentum components must be replaced with the summation over the momentum eigenvalues that form a discrete spectrum.
Since torsion increases with the magnitude of the four-momentum, the separation between adjacent eigenvalues also increases.
Consequently, the sum over the momentum eigenvalues converges.
We derived a prescription for this summation that gives a correct continuous limit for convergent integrals.
We extended this prescription to tensor integrals.
We showed that ultraviolet-divergent integrals turn into convergent sums, naturally eliminating the ultraviolet divergence in loop diagrams at all orders.

We applied our prescription to vacuum polarization and derived a finite, gauge-invariant vacuum polarization tensor.
We derived a finite running coupling that agrees with the low-energy limit of the standard QED.
Finally, we found that the modification of the photon propagator arising from the loops involving all charged fermions give a finite value for the bare electric charge of a particle: it is approximately 1.22 times its measured, renormalized charge.
We estimate that including the weak bosons $W$ and $Z$ contributing to vacuum polarization may change this value of the bare charge by a few percent, assuming that the spin-torsion coupling for the weak bosons is on the same order as that for fermions.
Torsional regularization of the $W$ and $Z$ loops will be considered elsewhere.

These results show that the bare charge of an electron $e_0$ and the renormalization constant for the photon wave function are finite.
The exact determination of the bare charge should elucidate the origin of the fine-structure constant, which is one of the unsolved problems in physics.
Consequently, including spin-torsion coupling naturally provides a physical regularization of ultraviolet divergences and makes renormalization paradigm finite and thus mathematically self-consistent.
Torsion may therefore be the physical source of a realistic regularization in QED (and in quantum field theory in general), eliminating the necessity of using auxiliary particles or varying the number of dimensions.
A more extensive analysis of torsional regularization and determination of the constant $U$ may use the quantum commutation relations between the metric and torsion tensors, derived from the Schwinger variational principle applied to the Einstein--Cartan action for the gravitational field \cite{rel}.

\begin{acknowledgements}
I am grateful to my parents Bo\.{z}enna Pop{\l}awska and Janusz Pop{\l}awski for their support, and to Gabe Unger for inspiring my research.
This work was funded by the University Research Scholar program at the University of New Haven.
\end{acknowledgements}

\end{document}